\documentclass[a4paper,runningheads]{llncs}
\usepackage[T1]{fontenc}
\usepackage{hyperref}
\usepackage{listings}
\usepackage{xcolor}
\usepackage{graphicx}
\usepackage{vdmlisting}
\usepackage{multirow}
\usepackage{array}
\usepackage{etoolbox} 
\makeatletter
\patchcmd{\ps@headings}{\rlap{\thepage}}{}{}{}
\patchcmd{\ps@headings}{\llap{\thepage}}{}{}{}
\makeatother
\newcommand\YAMLcolonstyle{\color{red}\mdseries}
\newcommand\YAMLkeystyle{\color{black}\bfseries}
\newcommand\YAMLvaluestyle{\color{blue}\slshape}

\makeatletter
\newcommand\language@yaml{yaml}

\expandafter\expandafter\expandafter\lstdefinelanguage
\expandafter{\language@yaml}
{
  keywords={true,false,null,y,n},
  keywordstyle=\color{darkgray}\bfseries,
  basicstyle=\YAMLkeystyle,                                 
  sensitive=false,
  comment=[l]{\#},
  morecomment=[s]{/*}{*/},
  commentstyle=\color{purple}\ttfamily,
  stringstyle=\YAMLvaluestyle\ttfamily,
  moredelim=[l][\color{orange}]{\&},
  moredelim=[l][\color{magenta}]{*},
  moredelim=**[il][\YAMLcolonstyle{:}\YAMLvaluestyle]{:},   
  morestring=[b]',
  morestring=[b]",
  literate =    {---}{{\ProcessThreeDashes}}3
                {>}{{\textcolor{red}\textgreater}}1     
                {|}{{\textcolor{red}\textbar}}1 
                {\ -\ }{{\mdseries\ -\ }}3,
  frame=lines,
}

\newcommand\ProcessThreeDashes{\llap{\color{cyan}\mdseries-{-}-}}

\lst@AddToHook{EveryLine}{\ifx\lst@language\language@yaml\YAMLkeystyle\fi}
\makeatother
\lstdefinestyle{VDM}
{
  frame=single,
  basicstyle=\small\ttfamily,
  escapechar=!,
  breaklines=true,
  frameround=false,
  linewidth=\columnwidth,
  morekeywords={atomic,is,inv,values,dcl,forall,in,set,nil,and,let,be,st,set1,pure,nat,pre,post,map,to,of,true,false},
  moredelim={[is][keywordstyle]{@}{@}},
}

\lstdefinestyle{TraceOutput}
{
  basicstyle=\small\ttfamily,
  frame=single,
  captionpos=b
}

\newcolumntype{R}[1]{>{\raggedleft\let\newline\\\arraybackslash\hspace{0pt}}m{#1}}

\begin{document}
\title{Cilium and VDM - Towards Formal Analysis of Cilium Policies}
%
%
\author{Tomas Kulik\inst{1} \and
Jalil Boudjadar\inst{2}}
\authorrunning{Kulik et al.}
%
\institute{Sweet Geeks \and Aarhus University, Denmark}
\maketitle              
\begin{abstract}
Industrial control systems are becoming more distributed and interconnected to allow for interaction with modern computing infrastructures. Furthermore, the amount of data generated by these systems is increasing due to integration of more sensors and the need to increase the reliability of the system based on predictive data models. One challenge in accommodating this data and interconnectivity increase is the change of the architecture of these systems from monolithic to component based, distributed systems. Questions such as how to deploy and operate such distributed system with many sub-components arise. One approach is the use of kubernetes to orchestrate the different components as containers. The critical nature of the industrial control systems however often requires strict component isolation and network segmentation to satisfy security requirements. Cilium is a popular network overlay for kubernetes that enables definition of network policies between different components running as kubernetes pods. The network policies are crucial for maintaining the secure operation of the system, however analysis of deployed policies is often lacking. In this paper, we explore the use of a formal analysis of Cilium network policies using VDM-SL. We provide examples of Cilium policies, an approach how they could be formalised using VDM-SL and analyse several scenarios to validate the policies against a model of simple real-life system.

\end{abstract}
\section{Introduction}
\label{sec:intro}
Industrial control systems are undergoing an architectural shift from isolated and often locally acting to interconnected and remotely accessible. While this trend has been ongoing for over a decade it has shown the need to secure this new architecture \cite{Nam2020,Kulik2019}. The system components are often seen as services that could be accessed remotely as well as initiate connections to other services. This allows the industrial control systems to be distributed and modular, allowing for addition of new services as the systems evolve. The challenges that arise from this approach span multiple areas, the distributed system must be robust, fault-tolerant, secure and safe. An approach to address the need for fault-tolerance and robustness could be the use of containerization and orchestration of containerized system components using kubernetes~\cite{poulton&2023} as adopted by Tesla, Amazon, Gemalto, etc. Deployment on kubernetes allows for scaling, load-balancing and replication of workloads allowing for increase in overall robustness of the system.

The security of such systems requires significant attention of its own. Significant amount of work has been done on the security of containers themselves~\cite{Brady&2020,Budigiri21,Flauzac&2020,Wenhao&2020}, however this covers the problematic of containerized workflows on kubernetes only partially. The need to manage and secure the internal network within a kubernetes system has led to development of many Container Network Interfaces (CNI) enabling for definition of network policies \cite{Nam2020}. This is important as the policy defines actions that were in the past often relegated to dedicated firewalls. The policies effectively instruct firewalls on the kubernetes nodes how to segment the network and what traffic flows shall be allowed within this network. One of the most popular CNIs is Cilium which utilizes networking based on extended Berkeley Packet Filter (eBPF) often providing performance and security improvements~\cite{Wustrich&2023}. The Cilium CNI utilized Cilium network policies created as kubernetes objects that are processed by the CNI within the deployed network. In the area of industrial control systems, these policies are used to control the flow of data such as commands, telemetry, alarms and potential configuration updates. These data-flows are not necessarily limited to sub-components running in kubernetes but also external components that either send data to or receive data from a service running in the kubernetes cluster. The primary task of securing the system using policies is precisely defining, which parts of the system can communicate and which parts shall be isolated \cite{Creane2021}. This leads to creation of specific corridors that the data can flow through. 

One challenge when utilizing the Cilium policies in a heavily distributed system consisting of many sub-components is how to ensure that the policy configuration is correct and follows the security requirements for the system \cite{Shamim2020}. To alleviate this challenge, a model-based approach with formal validation could be used. Formal methods have been utilized to provide strong assurances of cyber security properties within many areas including industrial control systems~\cite{Kulik&2022,Kulik2018}. One formal approach for modelling of Cilium policies using a formal language and utilizing a scenario based analysis of these policies is the use of VDM~\cite{Bjorner&2005}.

In this paper, we present a model-based analysis of Cilium policies using VDM-SL on a real-life industrial control system. We demonstrate the system model and apply formalized Cilium policies to this model. We model the policies in VDM based on the actual policies as deployed against a Cilium CNI in a kubernetes cluster. We then execute different data exchange scenarios within this system and analyze the impact of the policies on this data exchange. We create the model in a modular way allowing for addition or removal of policies between the different sub-components as well as addition or removal of different sub-components.

The rest of the paper is organized as follows: Section~\ref{sec:back} provides the necessary background on kubernetes objects, Cilium CNI and Cilium policies as well as VDM. Section~\ref{sec:arch} describes the architecture of the system used as a reference in this paper as well as introduces the data-flows within this architecture. Section~\ref{sec:pols} provides several network policies defined as a Cilium policy and shows how the policies apply to the system data-flows. Furthermore this section then demonstrates formalization of Cilium policies in VDM. Section~\ref{sec:analysis} describes the use of VDM for scenario-based analysis as well evaluates the results of the analysis. Section~\ref{sec:related} cites relevant related work and Section~\ref{sec:conclusion} presents conclusions and potential future work.


\section{Background}
\label{sec:back}
This section presents the background related to Kubernetes, Cilium and the VDM specification language.

\subsection{Kubernetes resources}
Kubernetes is a container orchestration system that utilizes an API based approach for a definition of different resources within a kubernetes cluster~\cite{poulton&2023}. The kubernetes cluster shall be understood as several individual machines (computers) running kubernetes, these machines are referred to as nodes. As kubernetes is a vast system, for the scope of this paper we consider the following resources: kubernetes pod, kubernetes namespace, kubernetes endpoints, kubernetes network policy and kubernetes deployment.

\textit{Kubernetes pod}: the pod resource is the smallest deployable computing unit that kubernetes can manage. Every pod can consist of a single or multiple containers. In case that the pod contains multiple containers, these are usually tightly coupled as the pod is often understood as running an application. It is therefore typical that pods often run a single container. For networking purposes, each pod obtains a unique IP address allocated from an internal kubernetes address space. In order for pods to be reachable from a network outside of the kubernetes cluster, they need to be configured with a forward facing service that obtains a specific IP address (different from a pod) but routable from the outside network. This IP address is often provided either be using the IP address of a kubernetes node or utilizing a load balancer subsystem that issues IP addresses from a predefined range.

\textit{Kubernetes namespace}: the namespace resource is a logical unit, often utilized to isolate sets of other resources (for example pods) from each other. Every namespace is assigned an internal DNS record so that the namespace could be used as a part of a DNS lookup. This is a typical case when applications running as pods need to communicate with a set of different applications running in separate namespaces. In the initial configuration, all pods within all namespace can communicate with all of the other pods as namespaces are primarily used as a resource name scoping and user segregation.

\textit{Kubernetes endpoints}: the endpoints resource is a collection of actual endpoints that are the communication points for a service. In its most basic form, the collection contains an IP address and a port on which the service will be discoverable. The endpoint can however contain multiple IP addresses and ports as well as information about a protocol, for example stating that it is a TCP endpoint, as well as its hostname.

\textit{Kubernetes network policy}: the network policy is an object understood by kubernetes and defines how different pods are allowed to communicate with each others. The policies define \textit{ingress} rules for the network traffic flowing into the pod or \textit{egress} rules for the network traffic flowing out of the pod. While kubernetes policies are primarily applied to pods, through use of selectors, they could be also applied against namespace or IP address blocks. The kubernetes network policies provide a basic set of rules for network segmentation primarily applicable to TCP and UDP protocols.

\subsection{Cilium and Cilium policies}
Cilium is an open source eBPF based CNI created to provide network connectivity to kubernetes workloads as well as secure and observe the network connectivity~\cite{Budigiri&2023}. The Cilium CNI contains several advanced features such as the address resolution protocol and the border gateway protocol routing support, load balancing, service mesh as well as security features of transparent encryption and runtime enforcement. Cilium provides its own network policy object named Cilium policy that significantly extends the functionality of basic kubernetes network policies.

Cilium policies provide policy definition on the layer 7 protocols of the OSI model~\cite{Day&83}, allowing to secure protocols such as HTTP, gRPC and many others. This in turn enables fine grained policy definitions without being limited to encapsulate all TCP or UDP traffic. Cilium policies also utilize identity based security, where each pod is assigned a unique security identity based on the labels (metadata) of the specific pod, decoupling the security from the IP address of the pod. The benefit of this is that the identities are based on immutable labels rather than IP addresses that can change.

\subsection{VDM Specification Language}
The Vienna Development Method specification language (VDM-SL) \cite{Bjorner&2005} is a ISO standardized formal language used for modeling and analyzing software systems. Developed initially at the IBM Vienna Laboratory in the 1970s, VDM provides a rigorous mathematical framework for specifying and validating the behavior of software components. The language supports both abstract and concrete specifications, allowing developers to describe system properties precisely and unambiguously. The language includes constructs for defining data types, functions and state transitions, facilitating comprehensive modeling of both functional and non-functional requirements. 

The systems behavior in VDM is specified as a set of functional units (modules or components), each of which is represented using the state-transition concept enabling constrained behavior using pre-conditions, post-conditions and invariants, referred to as \textit{contracts} \cite{Alagar1998}. A module may define a state component, which can be constrained by an invariant. A state component represents the valuation of a set of variables and parameters, where the enclosing module specifies the functionality to read and update the current values, in a constrained manner, leading thus to new states in the behavior. Functionality is defined by means of functions and operations over data types, where functions cannot read or alter the state directly while operations can. Consequently, functions are not permitted to call operations unless the operations are declared as \textit{pure}, meaning they are restricted from modifying the state but allowed to read it \cite{Tran19}.

For efficient testing and debugging, VDM-SL has been extended with traces concept to assist developers in identifying and refining behavior anomalies. With tool support for syntax checking, model validation and formal proof using the Visual Studio Code tools for VDM~\cite{Rask&22}, VDM remains a robust choice for developing high-assurance systems across various domains.

\section{System Architecture}
\label{sec:arch}
\begin{figure}
\centering
\includegraphics[width=1\textwidth]{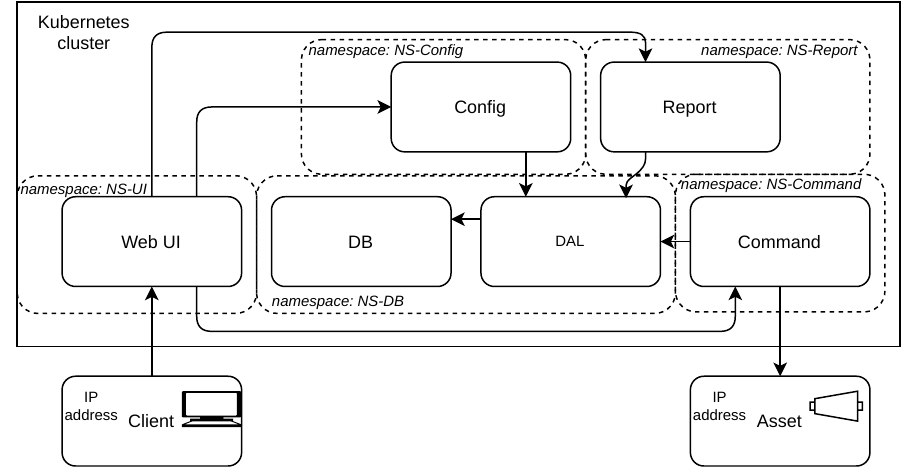}   
\caption{Containerized Industrial Control System}\label{fig:Architecture}
\end{figure}

The system we consider in this paper consists of an industrial control system deployed within kubernetes, accessible remotely by the operators controlling the physical infrastructure. The architecture is depicted in figure~\ref{fig:Architecture}. The \textit{Client} shall be understood as a computer utilized by the operators to access the industrial control system. Using the Client computer, the operator accesses the \textit{Web UI}, a web-based user interface that allows access to different features of the system. The features that the system allows is configuration of the system, for example setting up the information about the controlled assets, control of the assets and creation of reports about the functionality of the system. The system also contains a database in order to persist the configuration data and a log of commands that have been sent to the controlled asset. 

To achieve this functionality the system consist of several modules deployed as containers within Kubernetes. In the considered deployment, each container is deployed in a separate pod. The \textit{Config} module is responsible for loading and saving the system configuration data to the database. The \textit{Report} module is used to generate reports based on the operational data from the database, while the \textit{Command} module provides functionality to send commands to the controlled asset, while also storing the metadata about the commands to the database. All database operations are carried out by the data access layer (\textit{DAL}) module. The \textit{Database} itself is also containerized and deployed on Kubernetes. Finally, the system contains an \textit{Asset}, an external physical device that is controlled by the system. Each container is deployed in its own namespace with the exception of the \textit{Database} and \textit{DAL} that share a single namespace \textit{NS-DB} as they form a logical unit for storing and accessing data.

\section{Formalization of Cilium Network Policies}
\label{sec:pols}
The system we consider incorporates multiple data-flows. We specifically consider two system data-flows, both with external connectivity in order to capture most common use cases of an industrial control system. \textit{The first} data-flow considers the Client communication with the system deployed in Kubernetes. This covers the use case of an operator interacting with the system. \textit{The second} data-flow represents the use case of sending a command to the physical asset. This demonstrates the action of controlling an asset by a system from inside the Kubernetes cluster.

\textit{The Cilium policy} for the first data-flow defines the specific IP range given as a Classless Inter-Domain Routing (CIDR) block. The snippet of the policy covering the specific connectivity is shown in listing~\ref{lst:client_policy}. The policy shall be understood as the ingress where incoming connection is only allowed from a sub-net 10.28.1.2/30 as expressed in the \textit{fromCIDRSet} block, what leaves two usable IP addresses for an external client software (typically one active and one standby), the connections are only allowed to the port 443 of the \textit{WebUI} application as noted within the \textit{endpointSelector} block by use of the \textit{matchLabels} directive.

\begin{lstlisting}[language=yaml, caption={Cilium policy for the external client}, label={lst:client_policy}, float]
apiVersion: cilium.io/v2
kind: CiliumNetworkPolicy
metadata:
  name: UIPolicy
  namespace: NS-UI
spec:
  endpointSelector:
    matchLabels:
      app: WebUI
  ingress:
    - fromCIDRSet:
        - cidr: 10.28.1.2/30 #Client CIDR block
      toPorts:
        - ports:
            - port: "443"
\end{lstlisting}

In order to formalize this policy, we define several modules in VDM. We first create modules to represent the \textit{Application}. The application further utilizes a module representing the \textit{Endpoint}, which in turn consists of modules \textit{CIDR} and \textit{Namespace}. These modules are all created as vdmsl files in order to create logical structure for the model. The CIDR definition represents the four blocks of the IP address and the sub-net prefix length. The CIDR definition is shown in listing~\ref{lst:cidr}. Similarly, the namespace definition is a simple definition consisting of a record with two elements shown in listing~\ref{lst:ns}.

\begin{vdmsl}[style=VDM, caption={VDM CIDR definition}, label={lst:cidr}, float]
types
    FirstBlock = nat;
    SecondBlock = nat;
    ThirdBlock = nat;
    FourthBlock = nat;
    SigBits = nat;
CIDR::
    firstBlock : FirstBlock
    secondBlock : SecondBlock
    thirdBlock : ThirdBlock
    fourthBlock : FourthBlock
    sigBits : SigBits
\end{vdmsl}

\begin{vdmsl}[style=VDM, caption={VDM Namespace definition}, label={lst:ns}, float]
types
    Name = seq of char;
    Id = nat;
Namespace::
    name: Name
    id: Id
\end{vdmsl}

These definitions are then utilized within the \textit{Endpoint} definition as shown in listing~\ref{lst:ep}. The endpoint could use multiple identifiers such as CIDR, Namespace or label and additional connection information such as port.

\begin{vdmsl}[style=VDM, caption={VDM Endpoint definition}, label={lst:ep}, float]
types
  EndpointCidr = [CIDR];
  EndpointNamespace = [Namespace];
  EndpointPort = [nat];
  EndpointLabel = [seq of char];
Endpoint::
    endpointCidr : EndpointCidr
    endpointNamespace : EndpointNamespace
    endpointPort : EndpointPort
    endpointLabel : EndpointLabel
\end{vdmsl}

The \textit{Policy} is then modeled as a tuple of endpoints and a directional operator utilizing mapping. The directional operator states whether the endpoint tuple within the policy is an ingress or an egress policy. In the current model iteration, each policy only supports mapping between two endpoints. The policy model is shown in listing~\ref{lst:pol}.

\begin{vdmsl}[style=VDM, caption={VDM Policy definition}, label={lst:pol}, float]
types
  config = (Endpoint * Endpoint);
Policy::
    policy : map config to nat 
\end{vdmsl}

The application then models a deployed unit (internal or external) and can have several active endpoints on which it is listening for connections. The application model is shown in listing~\ref{lst:app}. It further defines an operation responsible for sending data to other applications within the system. The precondition on this operation checks whether this application is not only a sink application and can actually send data. The data is represented as a token type.

\begin{vdmsl}[style=VDM, caption={VDM Application definition}, label={lst:app}]
types
  AppId = nat;
  Endpoints = set of Endpoint;
  ReceiveOnly = bool;
  AppliedPolicies = set of Policy;
  SendEndpoint = Endpoint;

Application::
    sendEndpoint : SendEndpoint
    appID : AppId 
    activeListenEndpoints : Endpoints
    receiveOnly : ReceiveOnly
    appliedPolicies : AppliedPolicies

operations
SendData: AppId * AppId * Endpoint ==> ()
SendData(sid, rid, rep) ==
    TransferData(sid, rid, rep, mk_token(nil))
pre exists app in set applications & sid = app.appID 
    and app.receiveOnly = false;
\end{vdmsl}

Finally, the \textit{System} module holds the state of he overall system by keeping a record of all the applications deployed within the system as well as all the active endpoints and deployed policies. In order to build the system, the operator needs to deploy applications. This is modeled by operation \textit{DeployApplication} that utilizes the policies and endpoints deployed by the use of other operations. A partial content of the System module is shown in listing~\ref{lst:sys}.  

\begin{vdmsl}[style=VDM, caption={VDM System state and actions definition}, label={lst:sys}]
state SystemSt of
    applications : Applications
    policies: Policies
    endpoints: AllEndpoints
    appData: AppData
init s == s = mk_SystemSt({},{},{},{|->})
end    

types
  Applications = set of Application;
  Policies = set of Policy;
  AllEndpoints = set of Endpoint;
  AppData = map AppId to seq of token

operations

pure GetApplication: AppId ==> Application
GetApplication(aid) ==
    let app in set applications be st app.appID = aid
    in
    return app;

DeployApplication: AppId * Endpoint * set of Endpoint * bool * set of Policy  ==> ()
DeployApplication(aid, sendEp, recEps, ro, pols) ==
    let app = mk_Application(sendEp, aid, recEps, ro, pols) 
        in (
            applications := applications union {app}
        )
pre forall a in set applications & a.appID <> aid
post exists a in set applications & a.appID = aid;

TransferData: AppId * AppId * Endpoint * token ==> ()
TransferData(sapp, rapp, rep, d) ==
    appData(rapp):=appData(rapp) ^ [d]
pre exists ep in set endpoints & ep = rep and 
exists p in set policies & (dom p.policy = {mk_(rep, GetApplication(sapp).sendEndpoint)} and rng p.policy = {0}) 
or (dom p.policy = {mk_(GetApplication(sapp).sendEndpoint, rep)} and rng p.policy = {1})
post card elems appData(rapp) = card elems appData~(rapp) + 1;

\end{vdmsl}

The \textit{TransferData} operation under the System module represents data transfer from one application to another on a system level. This could be utilized by any two applications existing within the system state. It holds a precondition ensuring that the endpoints used within the data transfer are a part of a policy that is deployed within the system. The precondition checks if either an ingress or an egress policy allows for this data transfer. The helper operations for system setup, such as creation of endpoints and creation of policies is shown in listing~\ref{lst:help}. These operations are primarily present to provide complete interpretation of system deployment and network policy setup.

\begin{vdmsl}[style=VDM, caption={Helper operations for system deployment}, label={lst:help}]
CreatePolicy: Endpoint * Endpoint * nat ==> Policy
CreatePolicy(inEp, outEp, io) ==
    let pol = mk_Policy({mk_(inEp, outEp) |-> io})
        in (
            policies := policies union {pol}; 
            return pol
        )
pre forall pol in set policies & pol.policy <> {mk_(inEp, outEp) |-> io}
post exists pol in set policies & pol.policy = {mk_(inEp, outEp)|-> io};

CreateEndpoint: CIDR * Namespace * nat * seq of char ==> Endpoint
CreateEndpoint(cidr, ns, pt, lbl) ==
    let ep = mk_Endpoint(cidr, ns, pt, lbl)
        in (
            endpoints := endpoints union {ep};
            return ep
        )
pre forall ep in set endpoints & ep <> mk_Endpoint(cidr, ns, pt, lbl)
post exists ep in set endpoints & ep = mk_Endpoint(cidr, ns, pt, lbl);

\end{vdmsl}

\textit{The second data-flow Cilium policy} is shown in listing~\ref{lst:sec}. This policy allows for communication of the Web UI towards the controlled physical asset, specifically from the command module towards the controlled asset as an egress policy and from the WebUI to the Command module as an ingress policy as well. For demonstration purposes, we consider the egress part of this policy covering the communication from the Command application, as specified under the \textit{app: Command}, to the CIDR block 10.29.1.23/28 on port 5443.

\begin{lstlisting}[language=yaml, caption={Cilium policy for the external commanding}, label={lst:sec}]
apiVersion: cilium.io/v2
kind: CiliumNetworkPolicy
metadata:
  name: Command-Policy
  namespace: NS-Command
spec:
  endpointSelector:
    matchLabels:
      app: Command
  ingress:
    - fromEndpoints:
        - matchLabels:
            app: WebUI
            io.kubernetes.pod.namespace: NS-UI
  egress:
    - toCIDRSet:
        - cidr: 10.29.1.23/28
      toPorts:
        - ports:
            - port: "5443"
\end{lstlisting}

While the model components could be utilized to formally express parts of the Cilium policies, it is important to note that the model in its current iteration has several limitations. The policy can only contain a single endpoint, routing is not modeled limiting the analysis to basic network policies where higher layer protocols are not modeled. Chaining of policies has also not been considered in this work, and as such larger scenarios with multiple components with interacting policies could only be expressed with significant effort. We however consider this work a necessary stepping stone for defining how Cilium policies could be formalized in VDM.

\section{Formal Analysis}
\label{sec:analysis}
The formal analysis we conducted considers the two previously described data-flows. In both cases, two scenarios has been created, one with communication allowed by the policy and a second one with a policy violation. For the \textit{first data-flow}, the analysis scenario is shown in listing~\ref{lst:sc1}. The first operation of the analysis sets up the two applications, the client and the WebUI while assigning the correct endpoints and policies. In the second operation, the client attempts to transfer data to an unauthorized endpoint.

\begin{vdmsl}[style=VDM, caption={First data-flow analysis}, label={lst:sc1}]
Scenario1: () ==> ()
Scenario1()==
(
    dcl ep1 : Endpoint := CreateEndpoint(mk_CIDR(10,28,1,2,30), mk_Namespace("-", 0), 0, "");
    dcl ep2 : Endpoint := CreateEndpoint(mk_CIDR(0,0,0,0,0), mk_Namespace("NS-UI",1), 443, "WebUI");
    let pol = CreatePolicy(ep2, ep1, 0)
    in  
    (
        DeployApplication(1, ep1, {}, false, {pol});
        DeployApplication(2, ep2, {ep2}, true, {pol});
        appData := appData munion {2 |-> []};
        SendData(1,2,ep2)
    )
);

Scenario1PolicyViolation: () ==> ()
Scenario1PolicyViolation()==
(
    dcl ep1 : Endpoint := CreateEndpoint(mk_CIDR(10,28,1,2,30), mk_Namespace("-", 0), 0, "");
    dcl ep2 : Endpoint := CreateEndpoint(mk_CIDR(0,0,0,0,0), mk_Namespace("NS-UI",1), 443, "WebUI");
    dcl ep3 : Endpoint := CreateEndpoint(mk_CIDR(10,28,1,4,30), mk_Namespace("-", 0), 0, "");
    let pol = CreatePolicy(ep3, ep1, 0)
    in  
    (
        DeployApplication(1, ep1, {}, false, {pol});
        DeployApplication(2, ep2, {ep2}, true, {pol});
        appData := appData munion {2 |-> []};
        SendData(1,2,ep2)
    )
);
\end{vdmsl}

\noindent The first operation succeeds while the second operation has caused an expected precondition violation on the \textit{TransferData} operation.

The second data-flow is similar in nature to the first one, however it considers an egress data path. The analysis of this scenario is shown in listing~\ref{lst:sc2}. Again, in the scenario operations, two components are setup in this case: the command and the asset; where the command sends data to the external asset. In the second operation within this scenario, the command tries to egress data to an unauthorized endpoint.

\begin{vdmsl}[style=VDM, caption={Second data-flow analysis}, label={lst:sc2}]
Scenario2: () ==> ()
Scenario2()==
(
    dcl ep1 : Endpoint := CreateEndpoint(mk_CIDR(10,29,1,23,28), mk_Namespace("-", 0), 5443, "");
    dcl ep2 : Endpoint := CreateEndpoint(mk_CIDR(0,0,0,0,0), mk_Namespace("NS-Command",1), 0, "Command");
    let pol = CreatePolicy(ep2, ep1, 1)
    in  
    (
        DeployApplication(1, ep1, {}, false, {pol});
        DeployApplication(2, ep2, {ep2}, false, {pol});
        appData := appData munion {1 |-> []};
        SendData(2,1,ep1)
    )
);

Scenario2PolicyViolation: () ==> ()
Scenario2PolicyViolation()==
(
    dcl ep1 : Endpoint := CreateEndpoint(mk_CIDR(10,29,1,23,28), mk_Namespace("-", 0), 5443, "");
    dcl ep2 : Endpoint := CreateEndpoint(mk_CIDR(0,0,0,0,0), mk_Namespace("NS-Command",1), 0, "Command");
    dcl ep3 : Endpoint := CreateEndpoint(mk_CIDR(0,0,0,0,0), mk_Namespace("NS-Web",1), 0, "WebUI");
    let pol = CreatePolicy(ep2, ep1, 1)
    in  
    (
        DeployApplication(1, ep1, {}, false, {pol});
        DeployApplication(2, ep2, {ep2}, false, {pol});
        appData := appData munion {1 |-> []};
        SendData(2,1,ep3)
    )
);
\end{vdmsl}

Similar to the first scenario, the first operation succeeds while the second one leads to a precondition violation. The scenario execution time was negligible and scenario specification time could be considered fast enough once the model building blocks have been created.

\section{Related Work}
\label{sec:related}
Security analysis of containers in Kubernetes is crucial for ensuring the robustness and integrity of containerized applications in a dynamic orchestration environment. Recently, due to the large scale adoption of Kubernetes and container networks in general, many studies have explored the security analysis and enhancement of industrial control system deployed via Kubernetes \cite{Budigiri21,Nam2020,Tien2019,Kulik&2022,Brady&2020,Wenhao&2020}. Mostly, such analysis focuses on investigating whether the behavior of the integrated applications and nodes complies to the security policies such as container isolation, access controls, and network policies.  

The authors of \cite{Tien2019} proposed \textit{Kubanomaly}, a neural network based anomaly detection tool for the Kubernetes platform. It relies on monitoring the behavior for Kubernetes containers and a neural network classification model to identify abnormal behavior patterns. Although achieving high performance in attacks recognition, Kubanomaly may result in a considerable overhead time. Moreover, it cannot provide an absolute guarantee as the accuracy depends on the training dataset.

In \cite{Nam2020}, the authors proposed \textit{Bastion}, high-performance security enforcement tool that relies on extending the container hosting platform with an container-aware communication sandbox. The extension is built upon different services enabling to monitor and isolate the inter-container traffic to prevent access from other peer containers. This approach demonstrated a security improvement, however it may result in expensive overhead due to the dynamic of applications in the containers network.  


The authors of \cite{Jian2017} presented a status-based namespaces inspection as a security function to detect anomalous processes and prevent escape behaviors in cloud dockers. The proposed security enhancement is a kernel inline function designed to detect the user calls that involve different kernel stack space names by looking at the processes history, kill the process and track the
malicious user behind the container.

The authors of \cite{kang2022} introduced \textit{Verikube}, an efficient verification tool for container networks. Verikube relies on a graph structure to represent policies, so that to reduce memory consumption and computation time for verification, and run the containers data through the graph to verify whether a policy is violated or not. Such a verification is carried out using the \href{https://yices.csl.sri.com/}{Yices2} SMT solver. 

Yifan \textit{et al} \cite{Yifan20} introduced a container network policy verification, for applications orchestrated by Kubernetes, using an incremental approach where only the configuration changes are subject to verification. It relies on the labeled access control to containers as a sparse reachability relationship verified at runtime using a bimatrix analysis to reduce the performance overhead of verification.

\section{Conclusion}
\label{sec:conclusion}
In this paper, we have investigated the use of VDM for security analysis of Cilium policies. While the work is ongoing, several cases could already be formalized and analyzed as VDM scenarios. This has shown the potential of VDM to play a significant role in the analysis of modern systems based on software defined networks. Namely, we have presented basic building blocks as VDM modules that could be utilized to express a basic Cilium policy. While we have demonstrated how a standalone analysis could be carried out, our future plans are to integrate the VDM analysis into the deployment pipeline, where the Cilium policies are analyzed against a set of predefined scenarios before they are deployed to a production system. We further plan to expand the modules presented in this paper to allow for analysis of higher layer protocols as well as simpler definition of policies, where a single policy can capture multiple endpoints and also consider both ingress and egress communication directions. Besides, we plan to utilize combinatorial testing feature of VDM in order to automatically create significant test coverage. One additional idea for improvement is the use of External Format Reader of VDMJ to automatically translate Cilium policies to VDM entities.
%
%
%
%
\bibliographystyle{splncs04}
\bibliography{references}

\begin{thebibliography}{10}
\providecommand{\url}[1]{\texttt{#1}}
\providecommand{\urlprefix}{URL }
\providecommand{\doi}[1]{https://doi.org/#1}

\bibitem{Alagar1998}
Alagar, V.S., Periyasamy, K.: Vienna Development Method (1998)

\bibitem{Bjorner&2005}
Bj{\o}rner, D.: The vienna development method (vdm) software specification \&
  program synthesis. In: Mathematical Studies of Information Processing:
  Proceedings of the International Conference Kyoto, Japan, August 23--26,
  1978. pp. 326--359. Springer (2005)

\bibitem{Brady&2020}
Brady, K., Moon, S., Nguyen, T., Coffman, J.: Docker container security in
  cloud computing. In: 2020 10th Annual Computing and Communication Workshop
  and Conference (CCWC) (2020)

\bibitem{Budigiri&2023}
Budigiri, G.: Secure and scalable policy management in cloud native networking.
  In: Proceedings of the 24th International Middleware Conference: Demos,
  Posters and Doctoral Symposium (2023)

\bibitem{Budigiri21}
Budigiri, G., Baumann, C., Mühlberg, J.T., Truyen, E., Joosen, W.: Network
  policies in kubernetes: Performance evaluation and security analysis. In:
  2021 Joint European Conference on Networks and Communications; and 6G Summit
  (EuCNC/6G Summit) (2021)

\bibitem{Creane2021}
Creane, B., Gupta, A.: Kubernetes Security and Observability. " O'Reilly Media,
  Inc." (2021)

\bibitem{Day&83}
Day, J., Zimmermann, H.: The {OSI} reference model. Proceedings of the IEEE
  \textbf{71}(12),  1334--1340 (1983)

\bibitem{Flauzac&2020}
Flauzac, O., Mauhourat, F., Nolot, F.: A review of native container security
  for running applications. Procedia Computer Science  \textbf{175},  157--164
  (2020), the 17th International Conference on Mobile Systems and Pervasive
  Computing (MobiSPC),The 15th International Conference on Future Networks and
  Communications (FNC),The 10th International Conference on Sustainable Energy
  Information Technology

\bibitem{Jian2017}
Jian, Z., Chen, L.: A defense method against docker escape attack (2017)

\bibitem{kang2022}
Kang, H., Shin, S.: Verikube: Automatic and efficient verification for
  container network policies. IEICE TRANSACTIONS on Information and Systems
  \textbf{105}(12) (2022)

\bibitem{Kulik&2022}
Kulik, T., Dongol, B., Larsen, P.G., Macedo, H.D., Schneider, S.,
  Tran-J\o{}rgensen, P.W.V., Woodcock, J.: A survey of practical formal methods
  for security. Form. Asp. Comput.  \textbf{34}(1) (2022)

\bibitem{Kulik2019}
Kulik, T., Tran-J{\o}rgensen, P.W.V., Boudjadar, J.: Formal security analysis
  of cloud-connected industrial control systems. In: Innovative Security
  Solutions for Information Technology and Communications (2019)

\bibitem{Kulik2018}
Kulik, T., Tran-Jørgensen, P.W., Boudjadar, J., Schultz, C.: A framework for
  threat-driven cyber security verification of iot systems. In: 2018 IEEE
  International Conference on Software Testing, Verification and Validation
  Workshops (ICSTW) (2018)

\bibitem{Yifan20}
Li, Y., Jia, C., Hu, X., Li, J.: Kano: Efficient container network policy
  verification. In: 2020 IEEE Symposium on High-Performance Interconnects
  (HOTI) (2020)

\bibitem{Nam2020}
Nam, J., Lee, S., Seo, H., Porras, P., Yegneswaran, V., Shin, S.:
  $\{$BASTION$\}$: A security enforcement network stack for container networks.
  In: 2020 USENIX Annual Technical Conference (USENIX ATC 20). pp. 81--95
  (2020)

\bibitem{poulton&2023}
Poulton, N.: The kubernetes book. NIGEL POULTON LTD (2023)

\bibitem{Rask&22}
Rask, J., Madsen, F., Battle, N., Macedo, H., Larsen, P.: Visual studio code
  vdm support. In: 18th International Overture Workshop (2021)

\bibitem{Shamim2020}
Shamim, M.S.I., Bhuiyan, F.A., Rahman, A.: Xi commandments of kubernetes
  security: A systematization of knowledge related to kubernetes security
  practices. 2020 IEEE Secure Development (SecDev) pp. 58--64 (2020)

\bibitem{Tien2019}
Tien, C.W., Huang, T.Y., Tien, C.W., Huang, T.C., Kuo, S.Y.: Kubanomaly:
  Anomaly detection for the docker orchestration platform with neural network
  approaches. Engineering reports  \textbf{1}(5) (2019)

\bibitem{Tran19}
Tran-J\o{}rgensen, P.W.V., Kulik, T., Boudjadar, J., Larsen, P.G.: Security
  analysis of cloud-connected industrial control systems using combinatorial
  testing. In: Proceedings of the 17th ACM-IEEE International Conference on
  Formal Methods and Models for System Design (2019)

\bibitem{Wenhao&2020}
Wenhao, J., Zheng, L.: Vulnerability analysis and security research of docker
  container. In: 2020 IEEE 3rd International Conference on Information Systems
  and Computer Aided Education (ICISCAE) (2020)

\bibitem{Wustrich&2023}
W\"{u}strich, L., Schacherbauer, M., Budeus, M., Freiherr~von K\"{u}n\ss{}berg,
  D., Gallenm\"{u}ller, S., Pahl, M.O., Carle, G.: Network profiles for
  detecting application-characteristic behavior using linux ebpf. In: 1st
  Workshop on eBPF and Kernel Extensions eBPF '23 (2023)

\end{thebibliography}

\end{document}